\documentclass[12pt,twoside]{article}

\date{} 

\begin{document} 

\centerline{\bf ADVANCED STUDIES IN THEORETICAL PHYSICS, Vol. x, 200x, no. xx, xxx - xxx} 

\centerline{} 

\centerline{} 

\centerline {\Large{\bf A look on the internal structure of Teleparallel Gravity}} 

\centerline{} 

%\centerline{\Large{\bf }} 

\centerline{} 

\centerline{\bf {L.R.A. Belo*, E.P. Spaniol**, J.A. de Deus*** and V.C. de Andrade****}} 

\centerline{} 

\centerline{Institute of Physics, University of Brasilia 70.917-910, Brasilia, Distrito Federal, Brazil} 

\centerline{* leandrobelo@fis.unb.br} 

\centerline{** spaniol@fis.unb.br} 

\centerline{*** julianoalves@fis.unb.br} 

\centerline{**** andrade@fis.unb.br}

\centerline{} 

\newtheorem{Theorem}{\quad Theorem}[section] 

\newtheorem{Definition}[Theorem]{\quad Definition} 

\newtheorem{Corollary}[Theorem]{\quad Corollary} 

\newtheorem{Lemma}[Theorem]{\quad Lemma} 

\newtheorem{Example}[Theorem]{\quad Example} 

\begin{abstract}
Using the Dirac procedure to treat constraints dynamical sistems applied to gravitation, as described in the context of Teleparallel Equivalent of General Relativity (TEGR), we investigate, from the first class constraints, the gauge transformations in the fundamental field: the components of tetrads. We have shown that there is no an isotropy in physical space with respect to gauge transformations, i.e., given an arbitrary gravitational field, coming from a gauge transformation in the internal space, physical space reacts differently in the spatial and temporal components. By making an appropriate choice, we have found a gauge transformation for the components of tetrad field that allows a direct analogy with the gauge transformations of the Yang-Mills theory. In addition, to the flat case in which the algebra index is fixed, we get transformations similar those of the Electromagnetism. Moreover, still considering the flat case, the dependence of the gauge transformation parameter in the space-time variables is periodic, just like in the Electromagnetism. Furthermore, the gravitoelectric and gravitomagnetic fields as have recently been defined, make sense since they allow a direct relation with the momenta, which is analogous to what occurs in other gauge theories.
\end{abstract} 


\section{Introduction} An alternative description of General Relativity (GR) is the TEGR.
The usual formulation of TEGR in the literature is its Lagrangian version
\cite{hayashi1,hayashi2,aldrovandi3,andrade4,arcos5,aldrovandi6}; such a formulation, as well as the theory of Yang-Mills and Electromagnetism has a configuration space larger than necessary, resulting the appearance of constraints in its Hamiltonian formulation \cite{maluf7,chee8}, or in the Riemann-Cartan geometry, with local SO(3,1) symmetry \cite{blagojevic9,blagojevic10,blagojevic11}. In the case of Yang-Mills and Electromagnetism, the Hamiltonian formulation can be achieved by following the Dirac algorithm \cite{dirac11} to deal with constraint Hamiltonian systems. This same algorithm allows us to conclude that the first class constraints, primary or secondary, act as generators of gauge transformation, giving an interpretation for the extra degrees of freedom coming from the Hamiltonian formulation.

The equivalence principle states that the special relativity equations must be recovered in a locally inertial coordinate system, where the effects of gravitation are absent. Thus, based on this principle would be natural to expect that gravitation had a local Poincar\'e symmetry and that it was possible to describe it as a genuine gauge theory for this group. In fact, it is possible \cite{blagojevic9}. However, there are theoretical and experimental evidence, that from the most fundamental point of view (beyond the standard model) the Lorentz symmetry is broken \cite{Kostelecky12,Carroll13,Colladay14,Belich15,Songaila16,Moffat17,Coleman18}. This would eliminate the Poincar\'e group as the local symmetry group of gravity, leaving room only for the translational sector (or more general other).

In this paper, we show that starting from the TEGR Lagrangian contained in the literature it can be conclude that there is no an isotropy in physical space with respect to gauge transformations, i.e., given an arbitrary gravitational field, coming from a gauge transformation in the internal space, physical space reacts differently in the spatial and temporal components. In addition, to the flat case in which the algebra index is fixed, we get transformations similar those of the electromagnetism. We will also see that for the flat case, the dependence of the gauge transformation parameter in the space-time variables is periodic, as well in the Electromagnetism. Furthermore, we will show that the gravitoelectric and gravitomagnetic fields as have recently been defined \cite{spaniol19}, make sense since they allow a direct relation with the momenta, which is analogous to what occurs in other gauge theories.

\section{Momenta canonically conjugated to the tetrad field} 

As we will see in the next section, by introducing some new fields it is possible to find a consistent Hamiltonian description for the TEGR. In this description it will be possible, for example, to identify all the constraints of the theory. Let us see now the reasons why it will be necessary to follow this alternative procedure, i.e., why we simply could not straightly follow the Legendre transformation procedure.

The Lagrangian density associated with TEGR has the form \footnote{A tetrad field
$h_{a}=h_{a}{}^{\mu}\partial_{\mu}$ is a linear basis that relates
the metric $g$ to the metric of the tangent space $g=g_{a
b}dx^{a}dx^{b}$ by $g_{a b}=g_{\mu \nu}h_{a}{}^{\mu}h_{b}{}^{\nu}$.} \footnote{The torsion is defined by $T^\rho{}_{\mu \nu}
\;\equiv\Gamma^\rho{}_{\nu \mu} \;-\Gamma^\rho{}_{\mu \nu}$. The
object $\Gamma^\rho{}_{\nu \mu}$ is the Weitzenbock connection
defined by $\Gamma^\rho{}_{\nu \mu} \;\equiv
h_{a}{}^{\rho}\partial_{\mu}h^{a}{}_{\nu}$.} \cite{aldrovandi3}:
\begin{equation}
{\cal L} = \frac{h}{2k^2} \left[\frac{1}{4} \; T^\rho{}_{\mu \nu}
T_\rho{}^{\mu \nu} + \frac{1}{2} \; T^\rho{}_{\mu \nu} \; T^{\nu
\mu}{}_\rho - T_{\rho \mu}{}^{\rho} \; T^{\nu \mu}{}_\nu \right] \label{2.1}
\end{equation}
with  $h=det(h^{a}{}_{\mu})$ and $k=\frac{8\pi G}{c^4}$. This
expression can be rewritten in a more elegant form to get:
\begin{equation}
{\cal L} = \frac{h}{8k^2} \left[ \frac{1}{4} \, T^{a}{}_{\mu \nu}
T^{b}{}_{\rho \lambda} \,  N_{a b}{}^{\nu \rho, \nu \lambda} \right], \label{2.2}
\end{equation}
with $N_{a b}{}^{\mu \rho, \nu \lambda}$ being the tensor
responsible for all possible contractions of indices, given by:

\begin{eqnarray}
N_{a b}{}^{\mu \rho, \nu \lambda} &=& \frac{1}{2} \eta_{ab} \left[g^{\mu \rho} \; g^{\nu \lambda}- g^{\mu \lambda} \; g^{\nu \rho}\right] + \frac{1}{2} h_{a}{}^{\rho} \left[ h_{b}{}^{\mu}\; g^{\nu
\lambda}-h_{b}{}^{\nu}\; g^{\mu \lambda}\right]\nonumber \\
&-&\frac{1}{2}h_{a}{}^{\lambda} [h_{b}{}^{\mu}\; g^{\nu\rho}-h_{b}{}^{\nu}\; g^{\mu \rho}] + h_{a}{}^{\mu} \left[h_{b}{}^{\lambda}\; g^{\nu \rho}-h_{b}{}^{\rho}\; g^{\mu \lambda}
\right]\nonumber \\
&-& h_{a}{}^{\nu} \left[ h_{b}{}^{\lambda}\; g^{\mu \rho}-h_{b}{}^{\rho}\; g^{\mu \lambda}\right]. \label{2.3}
\end{eqnarray}
Another way to write the Lagrangian density (\ref{2.1}), and perhaps the most fruitful of all is:
\begin{equation}
\mathcal{L}_{G}=\frac{h}{4k^2}\;S^{\rho \mu \nu }\;T_{\rho \mu \nu }, \label{2.4}
\end{equation}
where
\begin{equation}
S^{\rho \mu \nu }=-S^{\rho \nu \mu }\equiv {\frac{1}{2}}\left[
K^{\mu \nu \rho }-g^{\rho \nu }\;T^{\theta \mu }{}_{\theta }+g^{\rho
\mu }\;T^{\theta \nu }{}_{\theta }\right] \label{2.5}
\end{equation}
and $K^{\mu \nu \rho }$ is the contortion tensor given by
\begin{equation}
K^{\mu \nu \rho}= \frac{1}{2}T^{\nu \mu \rho}+\frac{1}{2}T^{\rho \mu \nu}-\frac{1}{2}T^{\mu \nu \rho}. \label{2.6}
\end{equation}

As we would expect, this is a quadratic Lagrangian density in the
field strength tensor \footnote{Torsion written in the tetrad basis
$T^{a}{}_{\mu \nu}=h^{a}{}_{\rho}T^{\rho}{}_{\mu
\nu}=\partial_{\nu}h^{a}{}_{\mu}-\partial_{\mu}h^{a}{}_{\nu}$.}.

Let us first define the momenta canonically conjugated to tetrads
$h_c{}^{\sigma}$, using the Lagrangian density (\ref{2.2})
\begin{eqnarray}
\Pi_c{}^{\sigma} \equiv \frac{\partial L}{\partial(\partial_{0}
h^{c}{}_{\sigma})}=\frac{h}{8k^2} C_{c b}{}^{\rho \sigma
\lambda}T^{b}{}_{\rho \lambda}, \label{2.7}
\end{eqnarray}
with,
\begin{equation}
C_{c b}{}^{\rho \sigma \lambda}=\left[ N_{c b}{}^{0 \rho, \sigma
\lambda}-N_{c b}{}^{\sigma \rho, 0 \lambda}+ N_{b c}{}^{\rho 0,
\lambda \sigma}-N_{b c}{}^{\rho \sigma, \lambda 0} \right]. \label{2.8}
\end{equation}
Repeating the calculation for the momentum using as a starting point
the Lagrangian density (\ref{2.4}), we get:
\begin{eqnarray}
\Pi_c{}^{\sigma} \equiv \frac{\partial L}{\partial(\partial_{0}
h^{c}{}_{\sigma})}=-\frac{h}{k^2} S_{c}{}^{\sigma 0}. \label{2.9}
\end{eqnarray}
Taking then the definitions (\ref{2.3}), (\ref{2.5}) and (\ref{2.8}), we see that
expressions (\ref{2.7}) and (\ref{2.9}) for the momentum are totally equivalent.
The fact that the superpotential $S^{a \mu \nu}=h^{a}{}_{\rho}S^{\rho
\mu \nu}$ appears explicitly in the expression of the Lagrangian
density (\ref{2.4}) makes it more useful than the definition (\ref{2.2}).

Seizing the opportunity, we would to use the latter definition of momentum to corroborate recent definitions regarding the gravito-electric and gravito-magnetic fields \cite{spaniol19}:
\begin{eqnarray}
E_{a}{}^{i}=S_{a}{}^{0 i}, \nonumber \\
\epsilon^{i j k}B_{a k}=S_{a}{}^{i j}. \label{2.10}
\end{eqnarray}
Comparing the definition (\ref{2.10}) for the gravito-electric field with the expression
(\ref{2.9}) for the momentum, we see that:
\begin{eqnarray}
\Pi_c{}^{i}=\frac{h}{k^2} E_{c}{}^{i}. \label{2.11}
\end{eqnarray}
This result clearly shows that the definitions for the
gravito-electric and gravito-magnetic fields proposed in Ref
\cite{spaniol19} lead us to a result completely analogous to what
occurs with the theories of Yang-Mills and the Electromagnetism in
which the momenta are also directly related to the "electric" fields
of those theories. The main difference here is that these fields are
related to the superpotential $S^{a \mu \nu}$, while the theories of
Yang-Mills and Electromagnetism are related to the field strength
tensor, $E_{a}{}^{i}=F_{a}{}^{0 i}$ and $E^{i}=F^{0 i}$, respectively.
This difference is justified by the fact that gravitation, unlike
the other interactions, presents a special property, soldering
\cite{kobayashi20}. This property is a consequence of the existence
of a tetrads field $h^{a}{}_{\mu}$, which acts as a link between the
bundle (inner space) and the manifold, so that algebra indices can
be transformed into space-time indices, implying a Lagrangian density
over a quadratic term in the field strength tensor.

The usual sequence from here would be to isolate the $\partial_{0}h^{c}{}_{\sigma}$ terms of velocity and get the Hamiltonian version of the TEGR but unfortunately this can not be made in a simple way. Again, this difficulty is related to the fact that the momentum be related with superpotential and not with
the field strength. Let us see, from the expression (\ref{2.9}) we can show, after a long but straightforward calculation, that:
\begin{eqnarray}
\Pi_{c}{}^{i}+O\left(h^{c}{}_{i},\vec{\nabla} h^{c}{}_{i}\right)&=&[\frac{1}{2}\left(h_{c}{}^{0}h_{a}{}^{0}g^{ij}+g_{ac}g^{ij}g^{00}+h_{c}{}^{j}h_{a}{}^{i}g^{00}-h_{c}{}^{0}h_{a}{}^{i}g^{0j}\right) \nonumber \\
&-&h_{c}{}^{0}h_{a}{}^{0}g^{ij}-h_{c}{}^{i}h_{a}{}^{j}g^{00}+h_{c}{}^{i}h_{a}{}^{0}g^{0j}]\partial_{0}h^{a}{}_{j}. \label{2.12}
\end{eqnarray}
This set of equations can be rewritten as:
\begin{equation}
P_c{}^{i}=K_{ca}{}^{ij}\partial_{0}h^{a}{}_{j}, \label{2.13}
\end{equation}
where we define the objects
\begin{equation}
P_c{}^{i}\equiv \Pi_{c}{}^{i}+O\left(h^{c}{}_{i},\vec{\nabla} h^{c}{}_{i}\right) \label{2.14}
\end{equation}
and
\begin{eqnarray}
K_{ca}{}^{ij}&\equiv&\frac{1}{2}\left(h_{c}{}^{0}h_{a}{}^{0}g^{ij}+g_{ac}g^{ij}g^{00}+h_{c}{}^{j}h_{a}{}^{i}g^{00}-h_{c}{}^{0}h_{a}{}^{i}g^{0j}\right) \nonumber \\
&-&h_{c}{}^{0}h_{a}{}^{0}g^{ij}-h_{c}{}^{i}h_{a}{}^{j}g^{00}+h_{c}{}^{i}h_{a}{}^{0}g^{0j}, \label{2.15}
\end{eqnarray}
from where we obtain that
\begin{equation}
\partial_{0}h^{a}{}_{j}=\left(K^{-1}\right)^{ca}{}_{ij}P_{c}{}^{i}. \label{2.16}
\end{equation}
That is, find a set of solutions to the system (\ref{2.12}) is equivalent to find the inverse $\left(K^{-1}\right)^{ca}{}_{ij}$; such task seems impossible even if we use algebraic manipulators. To circumvent this issue, we will adopt another possible procedure as described below.

\section{Constraints as generators of gauge transformation} 
Motivated by the theory of Yang-Mills and by Electromagnetism,
where the secondary constraints are directly related to the Gauss's
law, we can also make such a comparison for the case of gravitation
described by TEGR in order to have an alternative way in obtaining
the constraints of the theory. Before proceeding, we note that the
expression for the momenta (\ref{2.9}) gives us directly the primary
constraints of TEGR
\begin{eqnarray}
\Phi_{c}=\Pi_{c}{}^{0}=-\frac{h}{k^2}S_{c}{}^{0 0}. \label{3.1}
\end{eqnarray}
To test if the gravitational Gauss's law \cite{spaniol19} really represents the secondary constraints we make \footnote{The canonical hamiltonian density is given by ${\cal H}_{0} = \Pi_{c}{}^{i}\partial_{0}h^{c}{}_{i}-{\cal L}$.}:
\begin{eqnarray}
\frac{d\Phi_{c}}{dt}&=&\{ \Phi_{c},{\cal H}_{0} \} \nonumber \\
&=&\frac{\delta
\Phi_{c}{}^{0}}{\delta h^{a}{}_{\rho}} \frac{\delta {\cal H}_{0}}{\delta \Pi_{a}{}^{\rho}}-\frac{\delta
\Phi_{c}{}^{0}}{\delta \Pi_{a}{}^{\rho}} \frac{\delta {\cal H}_{0}}{\delta h^{a}{}_{\rho}} \nonumber \\
&=&-\frac{\partial{\cal H}_{0}}{\partial h^{c}{}_{0}}+\partial_{\lambda}\frac{\partial{\cal H}_{0}}{\partial (\partial_{\lambda}h^{c}{}_{0})} \nonumber \\
&=&\frac{\partial{\cal L}}{\partial h^{c}{}_{0}}+\partial_{0}\Pi_{c}{}^{0}-\partial_{\lambda}\frac{\partial{\cal L}}{\partial (\partial_{\lambda}h^{c}{}_{0})} \nonumber \\
&=&\chi_{c}. \label{3.2}
\end{eqnarray}
Here, $\chi_c$ is the gravitational Gauss's law given by:
\begin{equation}
\chi_{a}=\partial_{i}(hS_{a}{}^{0 i})-k^2(hj_{a}{}^{0})=k^2\partial_{i}(\Pi_{a}{}^{i})-k^2(hj_{a}{}^{0})=\partial_{i}(hE_{a}{}^{i})-k^2(hj_{a}{}^{0})=0, \label{3.3}
\end{equation}
with
\begin{equation}
j_{a}{}^{\rho}=\frac{\partial L}{\partial
h^{a}{}_{\rho}}=\frac{h_{a}{}^{\lambda}}{k^2} \left[ T^{c}{}_{\mu
\lambda}S_{c}{}^{\mu
\rho}-\frac{1}{4}\delta_{\lambda}{}^{\rho}T^{c}{}_{\mu
\nu}S_{c}{}^{\mu \nu } \right] \label{3.4}
\end{equation}
assuming the role of a "vacuum source". Or, in a way that makes clear the nonlinear character of gravity,
\begin{eqnarray}
\chi_{a}&=&\partial_{i}(hE_{a}{}^{i})+k^2 h [ H^{b c}{}_{a i
j}E_{b}{}^{i}E_{c}{}^{j}+T^{b c}{}_{a n i j}\varepsilon^{j n k}
E_{c}{}^{i}B_{b}{}^{k}+ g_{r i} h^{c}{}_{j} \varepsilon^{j r k}(
E_{c}{}^{i}B_{a}{}^{k} \nonumber \\
&-&1/2E_{a}{}^{i}B_{c}{}^{k}) + J^{c}{}_{i
j}E_{c}{}^{i}E_{a}{}^{j}+K^{b c}{}_{a r i j n}\varepsilon^{i j
k}\varepsilon^{n r t} B_{c}{}^{k}B_{b}{}^{t} ]=0, \label{3.5}
\end{eqnarray}
where objects $H^{b c}{}_{a i j}$, $T^{b c}{}_{a n i j}$ ,
$J^{c}{}_{i j}$ and  $K^{b c}{}_{a r i j n}$ are combinations of terms of tetrads \footnote {See Ref. \cite{spaniol19} for the complete expressions.}. The expression (\ref{3.5}) justifies the interpretation of
$j_{a}{}^{\rho}$ as a source of vacuum. Moreover, $j_{a}{}^{\rho}$
is the energy-momentum tensor of the gravitational field \cite{andrade21}. Notice that (\ref{3.5}) is equivalent to zero component of GR field equations written in terms of gravitoelectric and gravitomagnetic fields.

The full equivalence between the GR and TEGR takes place within the equations of motion derived from the Lagrangian (\ref{2.4}) and Einstein-Hilbert,
\begin{equation}
{\cal L}_{GR}=-\frac{\sqrt{-g}}{2 k^2} R. \label{3.6}
\end{equation}
It can be shown that \cite{aldrovandi3}:
\begin{equation}
{\cal L}_{GR}={\cal L}_{TEGR}+\partial_\mu(\frac{2h}{k^2}T^{\nu\mu}{}_{\nu}). \label{3.7}
\end{equation}
Although the divergence term in the above expression does not contribute to the dynamics, represented by the Euler-Lagrange equations, we would think that this term has some relevance in the Hamiltonian formulation. That is not the case. Consider:
\begin{equation}
{\cal L}=\frac{h}{4k^2}\;S^{\rho \mu \nu }\;T_{\rho \mu \nu}\;+\partial_\mu(\frac{2h}{k^2}T^{\nu\mu}{}_{\nu}), \label{3.8}
\end{equation}
hence
\begin{eqnarray}
\Pi_c{}^{\sigma} \equiv \frac{\partial {\cal L}}{\partial(\partial_{0}
h^{c}{}_{\sigma})}=-\frac{h}{k^2} S_{c}{}^{\sigma 0}+f(h_{c}{}^{\sigma},\partial^{i}h_{c}{}^{\sigma}). \label{3.9}
\end{eqnarray}
From the previous lagrangian it may be noted again that there is no dependence on $\partial_{0}h^{a}{}_{0}$, so $\Pi_{c}{}^{0}\equiv\Phi_c$ remain primary constraints. Continuing:
\begin{eqnarray}
\frac{d\Phi_{c}}{dt}&=&\{ \Phi_{c},{\cal H}_{0} \} \nonumber \\
&.& \nonumber \\
&.& \nonumber \\
&.& \nonumber \\
&=&\chi_{c}. \label{3.10}
\end{eqnarray}
We can thus conclude that using the Lagrangean density (\ref{2.4}) is satisfactory and there is no information lost in the divergence term in (\ref{3.8}).

If we look closely, we see that the expression for the secondary constraints (gravitational Gauss's law) has an explicit dependence on velocities, and as we said previously, to isolate those terms depending on the momenta is not a trivial task. A method for obtaining such constraints has been developed in \cite{maluf7}; we will use here only the final result, since development is quite long. We have:
\begin{equation}
\chi_{c}=h_{c}{}^{0}{\cal H}_{0}+h_{c}{}^{i}F_{i}, \label{3.11}
\end{equation}
where
\begin{eqnarray}
{\cal H}_{0}&=&-h_{a 0}\partial_{k}\Pi^{a k}-\frac{kh}{4g^{0 0}}(g_{i k}g_{j l}P^{i j}P^{k l}-\frac{1}{2}P^2) \nonumber \\
&+&kh(\frac{1}{4}g^{i m}g^{n j}T^{a}{}_{m n}T_{a i j}+\frac{1}{2}g^{n j}T^{i}{}_{m n}T^{m}{}_{i j}-g^{i k}T^{j}{}_{j i}T^{n}{}_{n k}) \label{3.12}
\end{eqnarray}
and
\begin{equation}
F_{i}=h_{a i}\partial_{k}\Pi^{a k}-\Pi^{a k}T_{aki}+\Gamma^{m}T_{0mi}+\Gamma^{lm}T_{lmi}+\frac{1}{2g^{0 0}}(g_{i k}g_{j l}P^{k l}-\frac{1}{2}P)\Gamma^{j}. \label{3.13}
\end{equation}
Moreover, the objects were defined:
\begin{eqnarray}
P^{ik}&=&\frac{1}{2kh}(h_{c}{}^{i}\Pi^{ck}+h_{c}{}^{k}\Pi^{ci})+g^{0m}(g^{kj}T^{i}{}_{mj}+g^{ij}T^{k}{}_{mj}-2g^{ik}T^{j}{}_{mj}) \nonumber \\
&+&(g^{km}g^{0i}+g^{im}g^{0k})T^{j}{}_{mj}, \label{3.14}
\end{eqnarray}
\begin{equation}
\Gamma^{ik}=\frac{1}{2}(h_{c}{}^{i}\Pi^{ck}-h_{c}{}^{k}\Pi^{ci})-kh[-g^{im}g^{kj}T^{0}{}_{mj}+(g^{im}g^{0k}-g^{km}g^{0i})T^{j}{}_{mj}] \label{3.15}
\end{equation}
and
\begin{equation}
\Gamma^{k}=\Pi^{0k}+2kh(g^{kj}g^{0i}T^{0}{}_{ij}-g^{0k}g^{0i}T^{j}{}_{ij}+g^{00}g^{ik}T^{j}{}_{ij}). \label{3.16}
\end{equation}
The class test of the constraints shows that they are all first class \cite{maluf7}. Following the Dirac algorithm \cite{dirac11}, let us calculate then the transformations generated by the constraints, which do not modify the physical state of the system (gauge transformations):
\begin{eqnarray}
\delta h^{b}{}_{\rho}(x)&=&\int d^{3}x'
\left[\varepsilon_{1}^{a}(x')\{ h^{b}{}_{\rho}(x),\Phi_{a}(x)
\}+\varepsilon_{2}^{a}(x')\{ h^{b}{}_{\rho}(x),\chi_{a}(x) \} \right] \nonumber \\
&=&\int d^{3}x' \varepsilon_{1}^{a}(x') \left(\frac{\delta
h^{b}{}_{\rho}(x)}{\delta h^{c}{}_{\beta}(x')} \frac{\delta
\Phi_{a}(x)}{\delta \Pi_{c}{}^{\beta}(x')}-\frac{\delta
h^{b}{}_{\rho}(x)}{\delta \Pi_{c}{}^{\beta}(x')} \frac{\delta
\Phi_{a}(x)}{\delta h^{c}{}_{\beta}(x')} \right) \nonumber \\
&+&\int d^{3}x' \varepsilon_{2}^{a}(x') \left(\frac{\delta
h^{b}{}_{\rho}(x)}{\delta h^{c}{}_{\beta}(x')} \frac{\delta
\chi_{a}(x)}{\delta \Pi_{c}{}^{\beta}(x')}-\frac{\delta
h^{b}{}_{\rho}(x)}{\delta \Pi_{c}{}^{\beta}(x')} \frac{\delta
\chi_{a}(x)}{\delta h^{c}{}_{\beta}(x')} \right), \label{3.17}
\end{eqnarray}
that results in
\begin{equation}
\delta h^{b}{}_{\rho}=\delta^{0}_{\rho}\varepsilon^{b}_{1}+\nabla_{\rho}\varepsilon_{2}^{b}, \label{3.18}
\end{equation}
with
\begin{equation}
\nabla_{\rho}\varepsilon_{2}^{b}\equiv\delta_{\rho}^{i}\partial_{i}\varepsilon_{2}^{b}+\omega^{b}{}_{a \rho} \varepsilon_{2}^{a} \label{3.19}
\end{equation}
and
\begin{eqnarray}
\omega^{b}{}_{a \rho}&\equiv&-\frac{1}{g^{00}}\delta_{\rho}^{i}h_{a}{}^{0}h^{b}{}_{i}g^{0\mu}T^{j}{}_{j\mu}+\frac{3}{2g^{00}}\delta^{i}_{\rho}g^{0 b}h_{a i}g^{0\mu}T^{j}{}_{j\mu}+\frac{1}{2g^{00}}\delta^{i}_{\rho}h^{b 0}h_{a i}g^{0\mu}T^{j}{}_{j\mu} \nonumber \\
&+&\frac{3}{2}\delta^{i}_{\rho}h^{b}{}_{\mu}h_{a}{}^{\nu}T^{\mu}{}_{i \nu}+\delta^{i}_{\rho}g^{0 b}g_{0 \mu}h_{a}{}^{\nu}T^{\mu}{}_{i \nu}-\frac{1}{2}\delta^{i}_{\rho}g_{i \mu}h^{b \nu}h_{a}{}^{\alpha}T^{\mu}{}_{\nu \alpha} \nonumber \\
&+&\frac{1}{2}\delta^{i}_{\rho}h_{a i}h^{b \mu}T^{0}{}_{0 \mu}
-\delta^{i}_{\rho}h^{b}{}_{i}h_{a}{}^{\mu}T^{0}{}_{0 \mu}+\frac{1}{2}\delta^{i}_{\rho}\delta^{b}_{a}T^{0}{}_{0 i} \label{3.20}
\end{eqnarray}
playing the role of covariant derivative and connection, respectively. The connection that appears in the
definition (\ref{3.20}) is not a usual spin connection, as we would expect for the case of a covariant derivative acting on a 4-vector with internal index. The reason for this must be related to the fact that we are dealing with a theory in which indices of internal space can be taken into space-time indices. Continuing, we can write:
\begin{equation}
\delta h^{b}{}_{0}=\varepsilon^{b}_{1}, \label{3.21}
\end{equation}
and
\begin{equation}
\delta h^{b}{}_{i}=\nabla_{i}\varepsilon_{2}^{b}, \label{3.22}
\end{equation}
which makes possible claims that there is no an isotropy in physical space with respect to gauge transformations, i.e., given an arbitrary gravitational field, arising from a gauge transformation in the internal space, physical space reacts differently in the temporal and spatial components. In addition, for the flat case in which the algebra index is fixed, we get transformations similar to those of electromagnetism,
\begin{equation}
\delta h_{0}=\varepsilon_{1}, \label{3.23}
\end{equation}
and
\begin{equation}
\delta h_{i}=\partial_{i}\varepsilon_{2}. \label{3.24}
\end{equation}
This is analogous to that was shown in ref. \cite{spaniol19}, in which, in the weak field limit, the gravitational Maxwell's equations are analogous to the electromagnetism. Here we see that this analogy is still valid in a more fundamental context.

We can go ahead and rewrite (\ref{3.18}) as follows:
\begin{equation}
\delta h^{b}{}_{\rho}=\delta^{0}_{\rho}\varepsilon^{b}_{1}+\delta_{\rho}^{i}\partial_{i}\varepsilon_{2}^{b}+\omega^{b}{}_{a \rho} \varepsilon_{2}^{a}. \label{3.25}
\end{equation}
Introducing now the following relation between the parameters $\varepsilon^{b}_{1}$ and $\varepsilon^{b}_{2}$
\begin{equation}
\varepsilon^{b}_{1}=\partial_{0}\varepsilon_{2}^{b}, \label{3.26}
\end{equation}
we have:
\begin{equation}
\delta h^{b}{}_{\rho}=\partial_{\rho}\varepsilon_{2}^{b}+\omega^{b}{}_{a \rho} \varepsilon_{2}^{a}. \label{3.27}
\end{equation}
The subindex 2 can now be ignored,
\begin{equation}
\delta h^{b}{}_{\rho}=\partial_{\rho}\varepsilon^{b}+\omega^{b}{}_{a \rho} \varepsilon^{a}\equiv \nabla^{'}_{\rho}\varepsilon^{b}. \label{3.28}
\end{equation}
The above transformations allow a direct analogy with the gauge transformations obtained in the Yang-Mills theory.

It is importantly to stress out that the statement about the anisotropy of the physical space made earlier is still valid, since the transformations in which was based the statement did not take into account the hypothesis that relates the two transformation parameters.  

\section{Dependence of the gauge transformation parameter in the space-time variables}
In genuine gauge theories, it is possible to use an arbitrary gauge to find a differential equation for the gauge transformation parameter. For the case of the ETRG, however, to find one similar to the Lorenz gauge, for example, is not an easy task and what we can do is to use equations that are valid for construction. Consider the absolute parallelism condition \cite{aldrovandi3}:
\begin{equation}
D_{\nu}h^{b}{}_{\rho}=\partial_{\nu}h^{b}{}_{\rho}-\Gamma^{\alpha}_{\rho\nu}h^{b}{}_{\alpha}=0. \label{4.1}
\end{equation}
If we substitute the transformations (\ref{3.28}) in this equation, we get:
\begin{equation}
D_{\nu}\nabla^{'}_{\rho}\varepsilon^{b}=0, \label{4.2}
\end{equation}
in which $\nabla^{'}_\rho$ is the operator defined in (\ref{3.28}) and $D_{\nu}$ the usual covariant derivative used in (\ref{4.1}). Solving this set of differential equations in a general form, is a dificult task. Fortunately, in the flat limit we have:
\begin{equation}
\partial_{\nu}\partial_{\rho}\varepsilon^{b}=0. \label{4.3}
\end{equation}
The simplified form (\ref{4.3}) has the same shape for each index $b$. The equation can be rewritten as follows
\begin{equation}
\partial_{\nu}\partial_{\rho}\varepsilon=0, \label{4.4}
\end{equation}
or raising the first index in order to obtain a wave equation
\begin{equation}
\partial^{\rho}\partial_{\rho}\varepsilon=0. \label{4.5}
\end{equation}
Obviously, the solution of the previous equation is a plane wave. Thus, in the flat limit, the analogy with electromagnetism is completed. It is important to stress out that in Electromagnetism, the transformation parameter has a clear interpretation: a phase that calibrates the wave function \footnote{Making use of a quantum term.} of the source field in internal space. In the case of gravitation, even if we are at the flat limit, this interpretation is lost, once we have four parameters on (\ref{4.3}).

\section{Final remarks}
Starting from the TEGR Lagrangian contained in the literature, it is not possible to obtain a Hamiltonian formulation for this theory simply following the standard procedure of a Legendre transformation. The main reason for this impossibility is the momenta be associated with the superpotencial and not with the field strength tensor, being this fact, in turn, a consequence of soldering property which requires that the Lagrangian has more than one quadratic term in torsion. It was also shown that the gravitational Gauss' law is exactly the secondary constraints of the theory as usual occurs in Electromagnetism. Moreover, divergence term necessary to ensure equality between the TEGR and Einstein-Hilbert Lagrangian densities does not influence on the final results of the secondary constraints.

When we act with the first class constraints on the components of the tetrad field we obtained as second kind gauge transformations a similar structure of the Yang-Mills transformations, wich therefore generalizes the transformations of Electromagnetism. The reason for these transformations are similar and not identical to those obtained in Yang-Mills, is due to the fact that the spin connection obtained for the case of TEGR be able to take algebra indices to physical space indices. Again, this characteristic must be associated with soldering property which allow an exchange between objects defined in the physical and internal spaces. Furthermore, we show that there is no isotropy in physical space with respect to gauge transformations, i.e., given a gauge transformation in the internal space, physical space reacts differently in the spatial and temporal components. In fact, for the flat case, in which the algebra index is fixed, we arrive at the transformations analogous to those of Electromagnetism. It is important to note that this analogy had been obtained through the gravitational Maxwell equations \cite{spaniol19}, however, here the analogy was made in a more fundamental level.

By replacing the second kind gauge transformations in the absolute parallelism condition, we have obtained a highly coupled system of 64 differential equations. Luckily, for the flat case, the system is substantially simplified, allowing even to say that the internal space has an isotropic structure, i.e., the solutions are independent of the index that characterizes this space. This new system is easily solved, and its solution allows us to know that the dependence of the gauge transformation parameters in the variables of physical space is periodic, i.e., the solution is a plane wave.

With regard to the Gravitoelectromagnetism, by getting the relation (\ref{2.11}), we hope to have contributed to corroborate the definitions (\ref{2.10}), since they lead to a relationship between the momenta and GE fields completely analogous to what happens in Electromagnetism and Yang-Mills theory. The definitions (\ref{2.10}) allow to rewrite the expression for the gravitational Lorentz force in terms of GE and GM fields, giving us an alternative way to get the gravitomagnetic "drag" so mentioned in the literature \cite{spaniol20}.

%{\bf ACKNOWLEDGEMENTS.} This is a text of acknowledgements. 

{\bf Received: Month xx, 200x}


\begin{thebibliography}{99} 

\bibitem{hayashi1}
K. Hayashi and T. Shirafuji, Phys. Rev. D \textbf{19}, 3524 (1979).

\bibitem{hayashi2}
K.Hayashi, Phys. Lett. B \textbf{69}, 441 (1977).

\bibitem{aldrovandi3}
R. Aldrovandi and J.G. Pereira, "An Introduction to Teleparallel Gravity", Institute of Theoretical Physics, UNESP, S\~ao Paulo, Brazil. "http://www.ift.unesp.br/gcg/tele.pdf".

\bibitem{andrade4}
V. C. de Andrade and J. G. Pereira, Phys. Rev. D \textbf{56}, 4689 (1997).

\bibitem{arcos5}
H. I. Arcos, V. C. de Andrade and J. G. Pereira, Int. J. Mod. Phys. D \textbf{13}, 807 (2004) [gr-qc/0403074].

\bibitem{aldrovandi6}
R. Aldrovandi, J. G. Pereira and K. H. Vu, Gen. Rel. Grav. \textbf{36}, 101 (2004) [gr-qc/0304106].

\bibitem{maluf7}
J. W. Maluf and J. F. da Rocha-Neto, Physical Review D \textbf{64}, 8 (2001).

\bibitem{chee8}
G. Y. Chee, Ye Zhang, Yongxin Guo, "Gravitational energy-momentum and the Hamiltonian formulation of the teleparallel gravity", [arXiv:gr-qc/0106053v1].

\bibitem{blagojevic9}
Milutin Blagojevic, "Gravitation and Gauge Symmetries", Institute of Physics Publishing (IoP), Bristol, 2002.

\bibitem{blagojevic10}
M.Blagojevic and I. A. Nikolic, Phys. Rev. D \textbf{62}, 024021 (2000).

\bibitem{blagojevic11}
M.Blagojevic and I. A. Nikolic, Phys. Rev. D \textbf{64}, 044010 (2001).

\bibitem{dirac11}
Paul A. M. Dirac, "Lectures on Quantum Mechanics", Dover Publications, INC; 1st edition (2001).

\bibitem{Kostelecky12}
V.A. Kostelecky and S. Samuel, Phys. Rev. D \textbf{39}, 683 (1989).

\bibitem{Carroll13}
S.M. Carroll, G.B. Field and R. Jackiw, Phys. Rev. D \textbf{41}, 1231 (1990).

\bibitem{Colladay14}
D. Colladay and V.A. Kosteleck´y, Phys. Rev. D \textbf{55}, 6760 (1997); D. Colladay and V.A. Kostelecky, Phys. Rev. D \textbf{58}, 116002 (1998); S.R. Coleman and S.L. Glashow, Phys. Rev. D \textbf{59}, 116008 (1999).

\bibitem{Belich15}
H. Belich, J.L. Boldo, L.P. Colatto, J.A. Helayel-Neto, A.L.M.A. Nogueira, Phys. Rev. D \textbf{68}, 065030 (2003); Nucl. Phys. B - Supp. \textbf{127}, 105 (2004).

\bibitem{Songaila16}
A. Songaila and L.L. Cowie, Nature \textbf{398}, 667 (1999); P.C.W. Davies, T.M. Davies and C.H. Lineweaver Nature \textbf{418}, 602 (2002); A. Songaila and L.L. Cowie, Nature \textbf{428}, 132 (2004).

\bibitem{Moffat17}
J.W. Moffat, Int. J. Mod. Phys. D \textbf{12}, 1279 (2003); O. Bertolami, Lecture Notes in Physics, \textbf{633}, 121-139 (2004).

\bibitem{Coleman18}
S.R. Coleman and S.L. Glashow, Phys. Rev. D \textbf{59}, 116008 (1999); V.A. Kosteleck´y and M. Mewes, Phys. Rev. Lett. \textbf{87}, 251304 (2001); Phys. Rev. D \textbf{66}, 056005 (2002).

\bibitem{spaniol19}
E. P. Spaniol, V. C. Andrade. International Journal of Modern Physics D \textbf{19}, 489-505 (2010).

\bibitem{kobayashi20}
S. Kobayashi and K. Nomizu, Foundations of Diferential Geometry, 2nd edition (Intersciense, New York, 1996).

\bibitem{andrade21}
V. C. de Andrade, L. C. T. Guillen and J. G. Pereira, Phys. Rev. Lett. \textbf{84}, 4533 (2000).

\bibitem{spaniol20}
E. P. Spaniol, L. R. A. Belo, J. A. de Deus, V. C. de Andrade. The role of observers in the measurement of the Teleparallel Gravitoelectromagnetic fields in different geometries. (2011) (Submitted). 

\end{thebibliography}
\end{document}